\pdfoutput=1
\documentclass[sigconf]{acmart}

\newsavebox\CBox
\def\textBF#1{\sbox\CBox{#1}\resizebox{\wd\CBox}{\ht\CBox}{\textbf{#1}}}

\usepackage{amsmath,amssymb,amsfonts}
\usepackage{algorithmic}
\usepackage{graphicx}
\usepackage{xcolor}
\usepackage{setspace}
\usepackage{booktabs} 
\usepackage{subfigure}
\usepackage{amsmath}
\usepackage{color}
\usepackage{amsmath}
\usepackage{amssymb}
\usepackage{mathrsfs}
\usepackage[normalem]{ulem}
\usepackage{enumitem}
\usepackage{multirow}
\usepackage{ulem}
\usepackage[nomargin,inline,marginclue,draft]{fixme}
\usepackage{balance}
\usepackage{verbatim}
\usepackage{diagbox}
\usepackage{changepage}
\usepackage{bm}
\usepackage{makecell}

\usepackage[medium,compact]{titlesec}

\definecolor{myred}{rgb}{1, 0, 0}
\definecolor{myblue}{rgb}{0, 0, 1}
\definecolor{myblack}{rgb}{1, 1, 1}

\newlength\savedwidth

\definecolor{mycolor}{rgb}{0,0,0}
\acmConference[SIGIR'20]{the 43rd International ACM SIGIR Conference on Research and Development 
in Information Retrieval}{July 25--30, 2020}{Xi'an, China}
\acmYear{2020}

\title{Bundle Recommendation with Graph Convolutional Networks}
\author{Jianxin Chang$^{1}$, Chen Gao$^{1}$, Xiangnan He$^{2}$, Yong Li$^{1}$, Depeng Jin$^{1}$}
\affiliation{\institution{$^{1}$Beijing National Research Center for Information Science and Technology (BNRist),\\
Department of Electronic Engineering, Tsinghua University}
  }
\affiliation{\institution{$^{2}$School of Information Science and Technology, University of Science and Technology of China}}
\email{liyong07@tsinghua.edu.cn}
\renewcommand{\shortauthors}{Chang~\textit{et al.}}

\begin{document}
\begin{spacing}{0.85} 

\vspace{-0.2cm}
\begin{abstract}
\vspace{-0.1cm}
Bundle recommendation aims to recommend a bundle of items for a user to consume as a whole.
Existing solutions integrate user-item interaction modeling into bundle recommendation
by sharing model parameters or learning in a multi-task manner,
which cannot explicitly model the affiliation between items and bundles, 
and fail to explore the decision-making when a user chooses bundles.
In this work, we propose a graph neural network model named BGCN 
(short for \textit{\textBF{B}undle \textBF{G}raph \textBF{C}onvolutional \textBF{N}etwork}) 
for bundle recommendation.
BGCN unifies user-item interaction, user-bundle interaction and bundle-item affiliation 
into a heterogeneous graph.
With item nodes as the bridge, graph convolutional propagation between user and bundle nodes 
makes the learned representations capture the item level semantics.
Through training based on hard-negative sampler, 
the user's fine-grained preferences for similar bundles are further distinguished.
Empirical results on two real-world datasets demonstrate the strong performance gains of BGCN,
which outperforms the state-of-the-art baselines by 10.77\% to 23.18\%. 
\end{abstract}
\end{spacing}
\maketitle
\begin{spacing}{0.85} 
\vspace{-0.2cm}
\section{Introduction}
\vspace{-0.1cm}
The prevalence of bundled items on e-commerce and content platforms makes 
bundle recommendation become an important task. 
It is not only able to avoid users' monotonous choices to improve their experience, 
but also increase business sales by expanding the order sizes.
Since a bundle is composed of multiple items, the attractiveness of a bundle depends on its items, 
and the attractiveness of items in a bundle affects each other.
Besides, users need to be satisfied with most items in the bundle,
which means that there is a sparser interaction between the user and the bundle.

Most existing works for bundle recommendation~\cite{EFM,liu2014recommending,pathak2017generating}
regard item and bundle recommendation as two separate tasks, 
and associate them by sharing model parameters. 
A recent study~\cite{DAM} proposed a multi-task framework that 
transfers the benefits of the item recommendation task to the bundle recommendation
to alleviate the scarcity of user-bundle interactions. 
Despite effectiveness, we argue that they suffer from three major limitations:
\begin{itemize}[leftmargin=*,partopsep=0pt,topsep=0pt]
  \item \textBF{Separated modeling of two affiliated entities.} 
    Parameters sharing does not explicitly model the relationship between users, items and bundles,
    and the combination of multi-task way 
    is difficult to balance the weights of the main task and auxiliary task.
	\item \textBF{Substitution of bundles is not considered.} 
    Existing works only consider the correlation between items in a bundle to enhance the item task. 
    However, the association between the bundles as the recommended target is even more critical.
  \item \textBF{Decision-making is ignored when users interact with bundles.} 
    At the item level, even though a user likes most items in a bundle,
    but may refuse the bundle because of the existence of one disliked item.
    At the bundle level, for two highly similar bundles,
    the key to the user's final selection is their non-overlapping parts.
\end{itemize}

To address these limitations, we propose a solution named 
\textit{\textBF{B}undle \textBF{G}raph \textBF{C}onvolutional \textBF{N}etwork} (BGCN).
Utilizing the strong power of graph neural networks
in learning from complicated topology and higher-order connectivity, 
our BGCN effectively incorporates item-awareness into bundle recommender as follows,
a) It unifies user nodes, item nodes and bundle nodes into a heterogeneous graph, 
propagating embeddings between users and bundles with items as the bridge. \\
b) The bundle-item-bundle meta-path is built on the graph to capture 
an alternative relationship between bundles. \\
c) Through training with hard-negative samples, 
the user's decision-making when choosing bundles is further explored.

\vspace{0.05cm}
To summarize, the main contributions of this work are as follows.
\begin{itemize}[leftmargin=*,partopsep=0pt,topsep=0pt]
    \item To the best of our knowledge, we are the first to propose a graph neural network model 
      to explicitly model complex relations between users, items and bundles 
      to solve the problem of bundle recommender. 

    \item 
      We develop embedding propagation at two levels on the constructed graph,
      which distinguish affiliation between item and bundle nodes 
      to learn the representation of users and bundles with item level information.
      Training with hard negatives further explores the fine-grained differences between bundles.
    
    \item Extensive experiments on two real-world datasets show that 
      our proposed method outperforms existing state-of-the-art baselines by 10.77\% to 23.18\%.

\end{itemize}

\begin{figure*}[!t]
\vspace{-0.6cm}
\begin{center}
\subfigure
{\includegraphics[height=4cm]{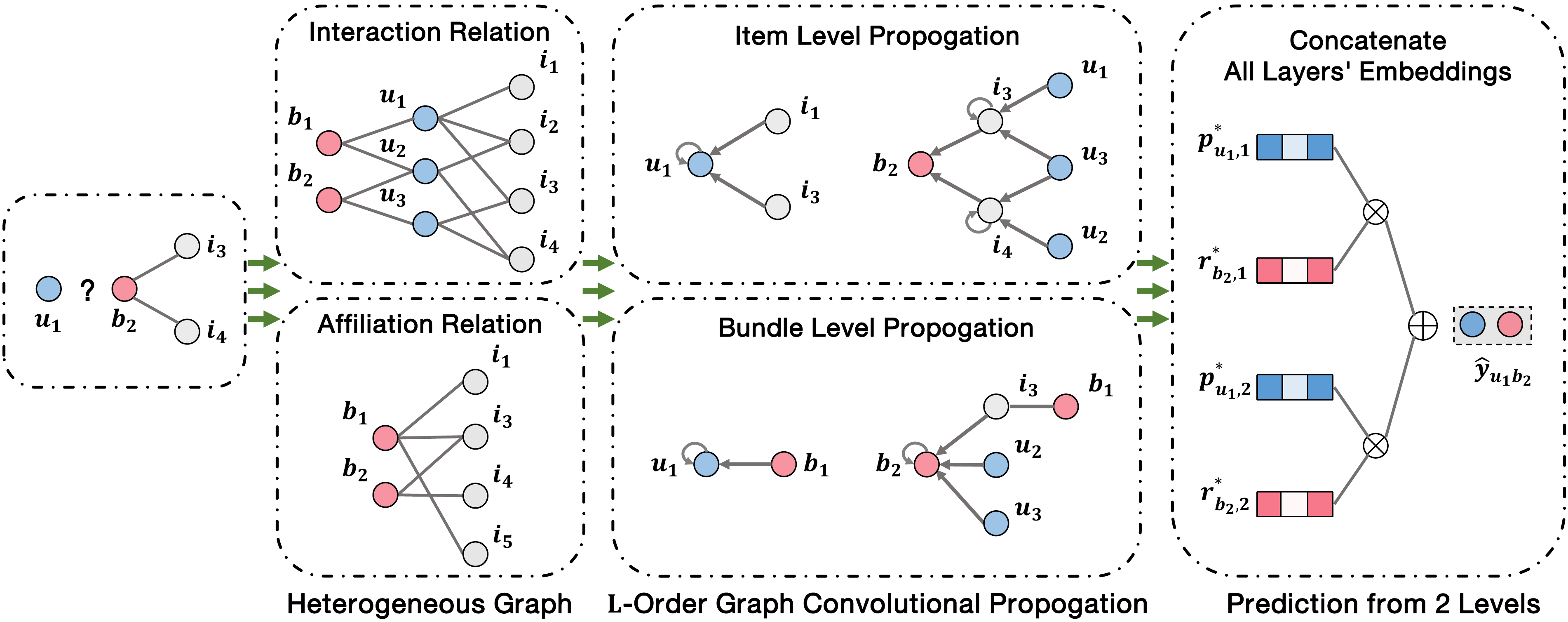}}
\hspace{-0.17cm}
\subfigure
{\includegraphics[height=4cm]{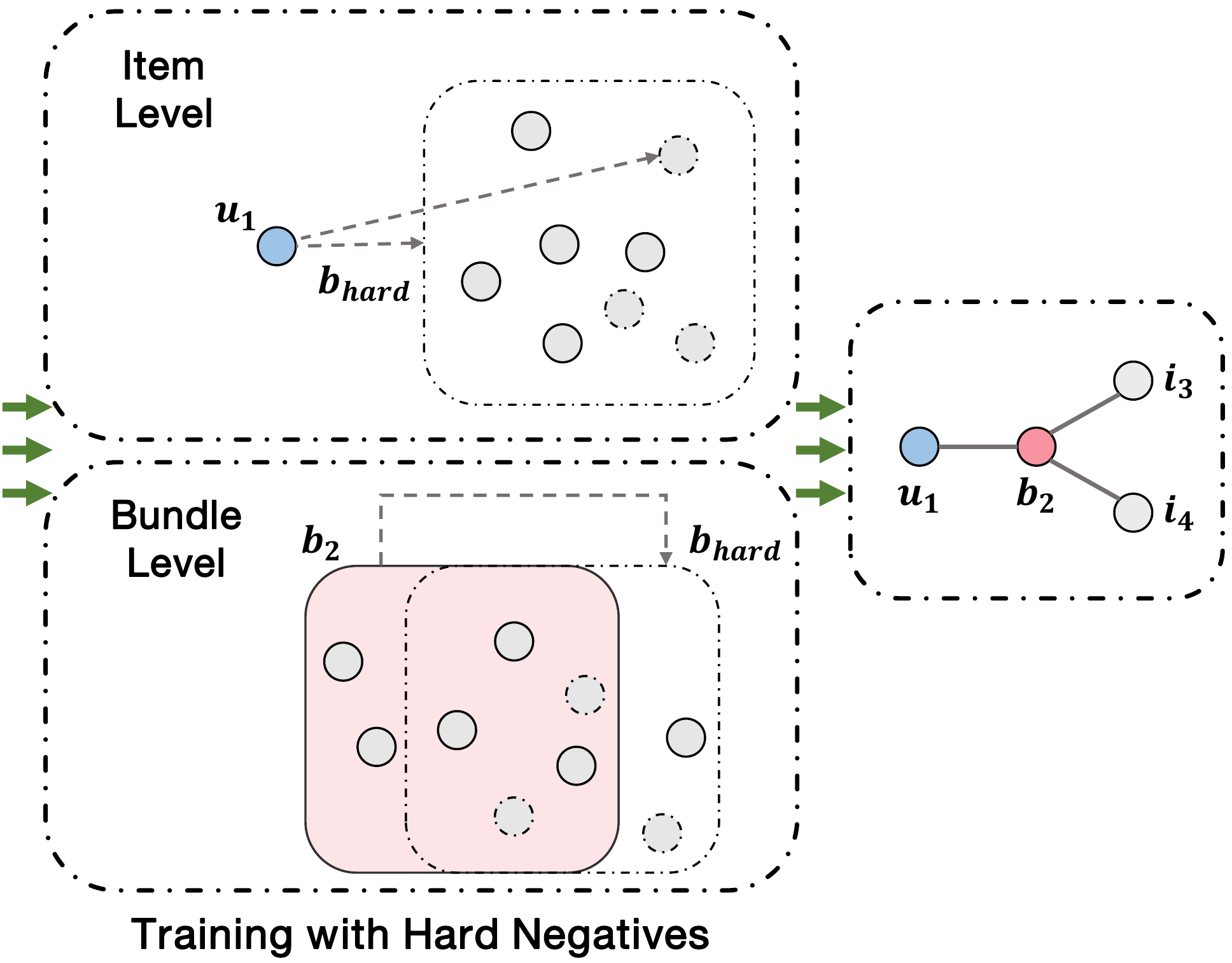}}
\end{center}
\vspace{-0.4cm}
\caption{The illustration of BGCN model, where $u_1$ is the target user and $b_2$ is the target bundle.
For a clear display, we divide the unfied heterogeneous graph into two kinds of 
relationships and draw them separately. Best view in color.
}\label{fig::framework}
\vspace{-0.4cm}
\end{figure*}

\vspace{-0.1cm}
\section{Problem Definition}
\vspace{-0.1cm}

In order to integrate item level information to improve bundle recommendation accuracy,
there are two types of important side information that need to be modeled, 
the user's preference to item and bundle's composition information.
We use $\mathcal{U}$, $\mathcal{B}$ and $\mathcal{I}$ to denote the set of users, bundles and items,
and define user-bundle interaction matrix, user-item interaction matrix,
and bundle-item affiliation matrix as
$ \mathbf{X}_{M \times N} =\{\mathbf{x}_{ub} | u \in \mathcal{U}, b \in \mathcal{B} \} $, 
$ \mathbf{Y}_{M \times O} =\{\mathbf{y}_{ui} | u \in \mathcal{U}, i \in \mathcal{I} \} $, and
$ \mathbf{Z}_{N \times O} =\{\mathbf{z}_{bi} | b \in \mathcal{B}, i \in \mathcal{I} \} $
with a binary value at each entry, respectively.
An observed interaction $\mathbf{x}_{ub}$ means user $u$ once interacted bundle $b$,
and an observed interaction $\mathbf{y}_{ui}$ means user $u$ once interacted item $i$.
Similarly, an entry $\mathbf{z}_{bi}=1$ means bundle $b$ contains item $i$. 
$M$, $N$ and $O$ denote the number of users, bundles and items, respectively.

Based on the above definition, the problem of bundle recommendation is then formulated as follows:

\textBF{Input:}  
user-bundle interaction data $ \mathbf{X}_{M \times N} $, 
user-item interaction data $ \mathbf{Y}_{M \times O} $,  
and bundle-item affiliation data $ \mathbf{Z}_{N \times O} $.

\textBF{Output:} 
A recommendation model that estimates the probability that user $u$ will interact with bundle $b$.

\section{Methodology}
\vspace{-0.1cm}

Figure~\ref{fig::framework} illustrates our proposed BGCN model
which is made up of the following three parts.

\begin{itemize}[leftmargin=*,partopsep=0pt,topsep=0pt]
  \item \textBF{Heterogeneous Graph Construction.}
We explicitly model the interaction and affiliation between users, 
bundles, and items by unifying them into a heterogeneous graph. 

  \item \textBF{Levels Propagation.}
The propagation at two levels on the constructed graph 
can capture the CF signal between users and items/bundles, 
the semantics of bundles and the alternatives between bundles
through distinguishing the affiliation relationship between item and bundle nodes.

  \item \textBF{Training with Hard Negatives.}
Considering the user's caution when choosing a bundle, 
the hard negatives further encode fine-grained representations of the user and the bundle.
\end{itemize}

\subsection{Heterogeneous Graph Construction}
\vspace{-0.1cm}
To explicitly model the relationship between users, bundles, and items,
we first build a unified heterogeneous graph. 
The interaction and affiliation data can be represented by an undirected graph
$\mathcal{G}=(\mathcal{V}, \mathcal{E} )$, 
where nodes are $\mathcal{V}$ consisting of user nodes $u \in \mathcal{U}$, 
bundle nodes $b \in \mathcal{B}$ and item nodes $i \in \mathcal{I}$,
and edges are $\mathcal{E}$ consisting of 
user-bundle interaction edges $(u,b)$ with $\mathbf{x}_{ub}=1$,
user-item interaction edges $(u,i)$ with $\mathbf{y}_{ui}=1$,
and bundle-item affiliation edges $(b,i)$ with $\mathbf{z}_{bi}=1$. 
The second block in Figure~\ref{fig::framework} illustrates our constructed graph.

For user, item and bundle nodes on the constructed graph, 
we apply one-hot encoding to encode the input
and compress them to dense real-value vectors as follows, 

\vspace{-0.2cm}
\begin{equation}\label{equation:embedding}
\small
	\begin{aligned}
    \mathbf{p}_u = \mathbf{P}^T\mathbf{v}_u^U, \quad
    \mathbf{q}_i = \mathbf{Q}^T\mathbf{v}_i^I, \quad
    \mathbf{r}_b = \mathbf{R}^T\mathbf{v}_b^B,
	\end{aligned}
\end{equation}
where $\mathbf{v}_u^U$, $\mathbf{v}_i^I$, $\mathbf{v}_b^B$ $ \in \mathbb{R}^N $
denotes the one-hot feature vector for user $u$, item $i$, and bundle $b$.
$\mathbf{P}$, $\mathbf{Q}$, and $\mathbf{R}$ denote the matrices of user embedding, item embedding, 
and bundle embedding, respectively.

\subsection{Item Level Propagation}
\vspace{-0.1cm}

The user's preference for an item in the bundle can attract the user's attention and interest to this bundle.
Since the bundled items are often carefully designed, they are often compatible with each other 
in function and compose some semantics to affect the user's selection context. 
For example, the bundle with a mattress and a bed frame reflects the meaning of bedroom furniture, 
and the bundle with a suit and a tie reflects the meaning of workplace dressing. 

To capture user preferences for the item and characteristic of the item itself,
we build upon an embedded propagation layer between user and item. 
Then information pooling from item to bundle can seize the 
semantic information of the bundle from the item level.
The propagation-based and pooling-based embedding updating rules 
for user $u$, item $i$ and bundle $b$ can be formulated as follows,

\vspace{-0.3cm}
\begin{equation}\label{eqn::item_level_propagation}
\small
\begin{aligned}
  & \mathbf{p}_{u,1}^{(\ell+1)} = 
  \sigma \left(W_{1}^{(\ell+1)} \left( \mathbf{p}_{u,1}^{(\ell)} + 
 {\textit{aggregate}} \left(\mathbf{q}_{i,1}^{(\ell)} | i\in \mathcal{N}_{u} \right)\right) + b_{1}^{(\ell+1)}\right), \\
  & \mathbf{q}_{i,1}^{(\ell+1)} = 
  \sigma \left(W_{1}^{(\ell+1)} \left( \mathbf{q}_{i,1}^{(\ell)} + 
  {\textit{aggregate}} \left(\mathbf{p}_{u,1}^{(\ell)} | u\in \mathcal{N}_{i} \right)\right) + b_{1}^{(\ell+1)}\right), \\
  & \mathbf{r}_{b,1}^{(\ell+1)} = 
  {\textit{aggregate}} \left(\mathbf{q}_{i,1}^{(\ell+1)} | i\in \mathcal{N}_{b} \right) ,\\
  & \mathbf{p}_{u,1}^{(0)} = \mathbf{p}_u, \mathbf{q}_{i,1}^{(0)} = \mathbf{q}_i,
\end{aligned}
\end{equation}
where $W_1$ is learned weight, $b_1$ is learned bias, 
$\sigma$ is non-linear activation function $\textit{LeakyReLU}$.
$\mathcal{N}_{u}, \mathcal{N}_{i}, \mathcal{N}_{b}$ represent 
neighbors of user $u$, item $i$ and bundle $b$, respectively.
We use the simple mean function as the aggregation function
and leave other more complex functions for future exploration.
Through the special propagation mechanism, the impact of bundle data sparsity can be weakened and
the cold start capability of the model may be improved naturally.

\subsection{Bundle Level Propagation}
\vspace{-0.1cm}
The close association between items in a bundle makes two bundles that share some items very similar.
The degree of similarity can be inferred from how many items they share. 
For example, computer sets that share five components are closer in performance than two,
and movie lists that share ten films are closer in theme than five.
For users, such bundles that share more items can often be considered at the same time.

We design a bundle-to-user embedding propagation module 
to learn preference for bundles from the bundle level. 
Then, a user-to-bundle embedding propagation is performed to extract bundle overall attributes.
Because highly overlapping bundles exhibit similar patterns in attracting users,
we develop weighted propagation based on the degree of overlap on the bundle-item-bundle meta-path
to seize alternative relationships between bundles.
The embedding updating rules at the bundle level can be formulated as follows,

\vspace{-0.20cm}
\begin{equation}\label{eqn::bundle_level_propagation}
\small
\begin{aligned}
  & \mathbf{p}_{u,2}^{(\ell+1)}= \sigma \left(W_{2}^{(\ell+1)} \left( \mathbf{p}_{u,2}^{(\ell)} + 
 {\textit{aggregate}} \left(\mathbf{r}_{b,2}^{(\ell)} | b\in \mathcal{N}_{u} \right)\right) + b_{2}^{(\ell+1)}\right), \\
  & \mathbf{r}_{b,2}^{(\ell+1)}= \sigma \left(W_{2}^{(\ell+1)} \left( \mathbf{r}_{b,2}^{(\ell)} + 
 {\textit{aggregate}} \left(\mathbf{p}_{u,2}^{(\ell)} | u\in \mathcal{N}_{b} \right)+ 
    \right. \right. \\
        & \ \ \ \ \ \ \ \ \ \ \ \ \ \ \ \ \ \ \ \ \   \left. \left. \
 {\textit{aggregate}} \left(\beta_{bb'} \cdot \mathbf{r}_{b',2}^{(\ell)} | b'\in \mathcal{M}_{b} \right) \right) +
 b_{2}^{(\ell+1)}\right), \\
        &\mathbf{p}_{u,2}^{(0)} = \mathbf{p}_u,  \ 
      \mathbf{r}_{b,2}^{(0)} = \mathbf{r}_b, \\
\end{aligned}
\end{equation}
where $W_2$ and $b_2$ are the trainable transformation matrix and bias, respectively.
$\mathcal{M}_{b}$ represents neighbors of bundle $b$ on the bundle-item-bundle meta-path.
$ \beta_{bb'} $ represents the overlap intensity between bundles after normalization.
Propagation of attributes expressed by similar $b$ helps bundles learn better representations 
and further enhance message-passing between $u$ and $b$.

\subsection{Prediction}
\vspace{-0.1cm}
After we iteratively do such propagation for $L$ times, we obtain $L$ user/bundle embeddings. 
We concatenate all layers' embeddings to combine the information 
received from neighbors of different depths for prediction. 
 
\vspace{-0.3cm}
\begin{equation}\label{eqn::concatenate}
\footnotesize
 \begin{aligned}
   &\mathbf{p}^{*}_{u,1} =  \mathbf{p}_{u,1}^{(0)} ||  \cdots ||  \mathbf{p}_{u,1}^{(L)}, 
   &\mathbf{r}^{*}_{b,1} =  \mathbf{r}_{b,1}^{(0)} ||  \cdots ||  \mathbf{r}_{b,1}^{(L)}, \\
   &\mathbf{p}^{*}_{u,2} =  \mathbf{p}_{u,2}^{(0)} ||  \cdots ||  \mathbf{p}_{u,2}^{(L)}, 
   &\mathbf{r}^{*}_{b,2} =  \mathbf{r}_{b,2}^{(0)} ||  \cdots ||  \mathbf{r}_{b,2}^{(L)}. 
     \end{aligned}
  \end{equation}
Finally, we use the inner product for final prediction 
and combine bundle and item levels as follows,

\vspace{-0.2cm}
\begin{equation}\label{eqn::prediction}
\small
    \begin{aligned}
\hat{y}_{ub} & = 
{\mathbf{p}_{u,1}^{*}}^{\top} \mathbf{r}_{b,1}^{*} + {\mathbf{p}_{u,2}^{*}}^{\top} \mathbf{r}_{b,2}^{*}.
     \end{aligned}
 \end{equation}	

\subsection{Training with Hard Negatives}
\vspace{-0.1cm}

Because bundles contain more items and have higher prices, 
users are often cautious when making decisions or spending money in bundle scenarios to avoid unnecessary risks.
For example, 
even though a user likes most items in a bundle, but may refuse the bundle because of the existence of one disliked item. 
For two highly similar bundles, the key to the user's final selection is their non-overlapping parts.

To optimize our BGCN model and take into account the user's decision-making when interacting with the bundle,
we design a learning strategy based on hard-negative samples.
Firstly, we adopt a pairwise learning manner that is widely used in implicit recommender systems~\cite{BPR}. 
Then after the model converges, the hard negative samples are introduced with a certain probability for more detailed training. 
Thus, we define the objective function as follows,
\begin{equation}
\footnotesize
  \begin{aligned}
  \mathrm{Loss} & = \sum_{(u,b,c) \in \mathcal{Q}}{-\mathrm\ln \sigma(\hat{y}_{ub}-\hat{y}_{uc})} + \lambda \cdot\|\Theta\|^{2},
  \end{aligned}
\end{equation}
where 
$\mathcal{Q} = \{(u,b,c)|(u,b)\in \mathcal{Y}^+, (u,c)\in \mathcal{Y}^- \}$ denote the pairwise training data with negative sampling. 
$\mathcal{Y}^+$ and $\mathcal{Y}^-$ denote the observed and unobserved user-bundle interaction, respectively. 
In the hard-negative sampler, for every $(u,b)\in \mathcal{Y}^+$, $c\in \mathcal{Y}^-$ is the bundle 
that $u$ has not interacted with but interacted with most of its internal items 
or that overlaps with $b$, as shown in the fifth block in Figure~\ref{fig::framework}.
To prevent over-fitting, we adopt L2 regularization 
where $\Theta$ stands for model parameters and $\lambda$ controls the penalty strength.

\vspace{-0.1cm}
\section{Experiments}
\vspace{-0.1cm}

In this section, we conduct extensive experiments
 to answer the following three research questions:
\begin{itemize}[leftmargin=*,partopsep=0pt,topsep=0pt]
\item \textBF{RQ1:} 
  How does our proposed BGCN model perform as compared with the state-of-the-art methods?
\item \textBF{RQ2:} 
  How do the key designs in our model affect performance?
\item \textBF{RQ3:} 
  How can BGCN alleviate the data sparsity problem in bundle recommender?
\end{itemize}

\vspace{-0.1cm}
\subsection{Experimental Settings}
\vspace{-0.1cm}
\subsubsection{Datasets and Metrics}
We use the following two real-world datasets for evaluation, with their statistics shown in Table~\ref{tab::datasets}.
\begin{itemize}[leftmargin=*,partopsep=0pt,topsep=0pt]
  \item \textBF{Netease}\footnote{https://music.163.com}
    This is a dataset from the largest music platform in China collected by~\cite{EFM}. 
    It enables users to bundle songs with a specific theme or add any bundles to their favorites.
  \item \textBF{Youshu}\footnote{http://www.yousuu.com}
    This dataset is constructed by~\cite{DAM} from Youshu, a Chinese book review site. 
    Similar to Netease, every bundle is a list of books that users desired.
\end{itemize}
We adopt two widely used metrics, Recall@K and NDCG@K, to judge the performance of the ranking list,
where Recall@K measures the ratio of test bundles that have been contained by the top-K ranking list,
while NDCG@K complements Recall by assigning higher scores to the hits at higher positions of the list.

\vspace{-0.1cm}
\subsubsection{Baselines and Parameters}
To demonstrate the effectiveness of our BGCN model, we compare it with the following six state-of-the-art methods. 
\begin{itemize}[leftmargin=*,partopsep=0pt,topsep=0pt]
	\item \textBF{MFBPR}\cite{BPR} 
    This is a matrix factorization method under a Bayesian Personalized Ranking pairwise learning framework, 
    which is widely used for implicit feedback.
	\item \textBF{GCN-BG}\cite{GCMC} 
    This is a widely used graph neural network model in the recommendation. 
    In the method, we apply GCN to the user-bundle bipartite interaction graph. 
	\item \textBF{GCN-TG}\cite{GCMC} 
    We use the same way with BGCN to build the user-item-bundle tripartite unified graph, 
    but the difference is that the message is passed between all kinds of nodes at the same time. 
	\item \textBF{NGCF-BG}\cite{NGCF}
   This is the state-of-the-art model which uses graph neural network to extract higher-order connectivity for the recommendation.
   Similar to GCN-BG, here only user and bundle are used to build the interaction bipartite graph.
	\item \textBF{NGCF-TG}\cite{NGCF}
   NGCF-TG performs propagation in the same way as GCN-TG on the tripartite graph. 
   Since item information is introduced, the embedding learned by the model might have stronger representation ability.
	\item \textBF{DAM}\cite{DAM}
   This is the state-of-the-art deep model in bundle recommendation.
   It uses the attention mechanism and multi-task learning framework to extend NCF\cite{he2017neural}.
\end{itemize}
For all these methods,
we adopt BPR loss and set the negative sampling rate to 1.
We employ Adam optimizer with the 2048-size mini-batch and fit the embedding size as 64.
The learning rate is searched in \{1e-5, 3e-5, 1e-4, 3e-4, 1e-3, 3e-3\}
and $L_2$ regularization term is tuned in \{1e-7, 1e-6, 1e-5, 1e-4, 1e-3, 1e-2\}.
For GCN-based methods, 
we search the message dropout and node dropout within  \{0, 0.1, 0.3, 0.5\}.
For training, 
the hard-negative samples are selected with an 80\% probability.

\begin{table}[t]
\vspace{-0.2cm}
\caption{Statistics of two utilized real-world datasets}\label{tab::datasets}
\vspace{-0.4cm}
\center
\footnotesize
\begin{tabular}{ccccccc}
\toprule
\textBF{Dataset}  
& \textBF{\#U}  & \textBF{\#I}  & \textBF{\#B}  & \textBF{\#U-I}  & \textBF{\#U-B}  & \textBF{\#Avg. I in B}  \\ 
\midrule
\textBF{Netease}  
& 18,528 & 123,628 & 22,864 & 1,128,065  & 302,303  & 77.80         \\ 
\textBF{Youshu}   
& 8,039  & 32,770  & 4,771  & 138,515    & 51,377   & 37.03         \\ 
\bottomrule  
\end{tabular}
\vspace{-0.5cm}
\end{table}

\begin{table*}[ht]
\footnotesize
\center
\vspace{-0.3cm}
\caption{Performance comparisons on two real-world datasets with six baselines} 
\label{tab::performance}
\vspace{-0.3cm}
\setlength{\tabcolsep}{1.2mm}{\begin{tabular}{lcccccccccccccc}
\toprule
\multirow{3}{*}{\textBF{Method}} & &\multicolumn{6}{c}{\textBF{Netease}} & &\multicolumn{6}{c}{\textBF{Youshu}} \\
\cmidrule(lr){3-8}  \cmidrule(lr){10-15} 
&
& \textBF{Recall@20}   & \textBF{NDCG@20}
& \textBF{Recall@40}   & \textBF{NDCG@40}
& \textBF{Recall@80}   & \textBF{NDCG@80}
&
& \textBF{Recall@20}   & \textBF{NDCG@20}
& \textBF{Recall@40}   & \textBF{NDCG@40}
& \textBF{Recall@80}   & \textBF{NDCG@80} \\
\midrule
MFBPR 
& & 0.0355          & 0.0181          & 0.0600          & 0.0246          & 0.0948          & 0.0323          &  & 0.1959          & 0.1117          & 0.2735          & 0.1320          & 0.3710          & 0.1543          \\
GCN-BG 
& & 0.0370          & 0.0189          & 0.0617          & 0.0255          & 0.1000          & 0.0342          &  & 0.1982          & 0.1141          & 0.2661          & 0.1322          & 0.3633          & 0.1541          \\
GCN-TG 
& & 0.0402          & 0.0204          & 0.0657          & 0.0272          & 0.1051          & 0.0362          &  & 0.2032          & 0.1175          & 0.2770          & 0.1371          & 0.3804          & 0.1605          \\
NGCF-BG 
& & 0.0395          & 0.0207          & 0.0646          & 0.0274          & 0.1021          & 0.0359          &  & 0.1985          & 0.1143          & 0.2658          & 0.1324          & 0.3542          & 0.1524          \\
NGCF-TG 
& & 0.0384          & 0.0198          & 0.0636          & 0.0266          & 0.1015          & 0.0350          &  & \underline{0.2119}    & 0.1165          & 0.2761          & 0.1343          & 0.3743          & 0.1561          \\
DAM
& & \underline{0.0411}    & \underline{0.0210}    & \underline{0.0690}    & \underline{0.0281}    & \underline{0.1090}    & \underline{0.0372}    &  & 0.2082          & \underline{0.1198}    & \underline{0.2890}    & \underline{0.1418}    & \underline{0.3915}    & \underline{0.1658}    \\
\textBF{BGCN}
& & \textBF{0.0491} & \textBF{0.0258} & \textBF{0.0829} & \textBF{0.0346} & \textBF{0.1304} & \textBF{0.0453} &  & \textBF{0.2347} & \textBF{0.1345} & \textBF{0.3248} & \textBF{0.1593} & \textBF{0.4355} & \textBF{0.1851} \\
\midrule
\% Improv.
& & 19.67\%         & 22.89\%         & 20.17\%         & 23.18\%         & 19.65\%         & 21.76\%         &  & 10.77\%         & 12.22\%         & 12.36\%         & 12.33\%         & 11.23\%         & 11.62\% \\
\bottomrule            
\end{tabular}
}
\vspace{-0.1cm}
\end{table*}

\subsection{Overal Performance (RQ1)}
\vspace{-0.1cm}
The experiment results are reported in Table~\ref{tab::performance}. 
From the results, we have the following observations.

\begin{itemize}[leftmargin=*]
    \item \textBF{Our proposed BGCN achieves the best performance.} 
      We can observe that our model BGCN significantly outperforms all baselines in terms of Recall and NDCG metrics.
      Specifically, BGCN outperforms the best baseline by 19.65\% - 23.18\% and 10.77\% - 12.36\% 
      on the Netease dataset and Youshu dataset, respectively. 
      Due to the special design, BGCN is the best model for utilizing graph structure and item interaction at the same time.
    
    \item 
      \textBF{Graph models are effective but not enough.} 
      Although the better performance of GCN and NGCF compared to MFBPR proves the strong power of the graph model, 
      they fail to exceed the state-of-the-art baseline DAM that leverage item interaction to recommend bundle. 
      Therefore, designs to make graph neural network work in bundle task is necessary, 
      which validates our motivation of incorporating items into the user-bundle graph.
 
    \item \textBF{More input does not always mean better performance.}
      The performance of GCN-TG and NGCF-TG with the same constructed graph as BGCN 
      is not always better than GCN-BG and NGCF-BG\@. 
      This shows that if the model itself cannot distinguish the affiliation relation 
      between item and bundle nodes, the extra item input may bring the noise. 
      Therefore, our special design for propagation in the tripartite unified graph is necessary.

\end{itemize}

\subsection{Ablation Study (RQ2)}
\vspace{-0.1cm}
To evaluate the effectiveness of several key designs in BGCN, 
we performed ablation studies as shown in Table~\ref{tab::ablation}.

\begin{itemize}[leftmargin=*]
    \item \textBF{Effectiveness of Levels Propagation} 
We compare the performance of the model performing propagation
at the only item level, only bundle level, and both levels.
The results show that the model with two levels outperforms the model with on only bundle level 
by 9.36\% - 11.57\% and 2.75\% - 11.39\% on the Netease dataset and Youshu dataset, respectively.
    \item \textBF{Effectiveness of B2B Propagation} 
The impact of overlap-based bundle propagation is tested by comparing the performance of models
with no b2b propagation, unweighted b2b propagation, and weighted b2b propagation.
The performance gains of the model introduced overlap-based b2b propagation 
are 5.84\% - 6.74\% and 1.79\% - 9.02\% on the Netease dataset and Youshu dataset, respectively.
It is shown that adding b2b propagation can help extract bundle correlations,
especially when the degree of overlap is injected into the model.
    \item \textBF{Effectiveness of Hard-negative sample} 
We compare the performance of the model under training with no hard samples,
and hard samples at the item level, the bundle level, and both levels, respectively.
We can observe that the model with hard-negative sample performs better than the model with simple
sample by 9.26\% - 10.64\% and 3.95\% - 11.78\% on the Netease dataset and Youshu dataset, respectively.
This demonstrates that hard-negative sample improves performance, 
especially when considering hard-negative samples at both item and bundle levels.
\end{itemize}

\subsection{Impact of Data Sparsity (RQ3)}
\begin{figure}[t]
\vspace{-0.3cm}
\centering
\hspace{-0.4cm}
\subfigure
{\includegraphics[height=2.7cm]{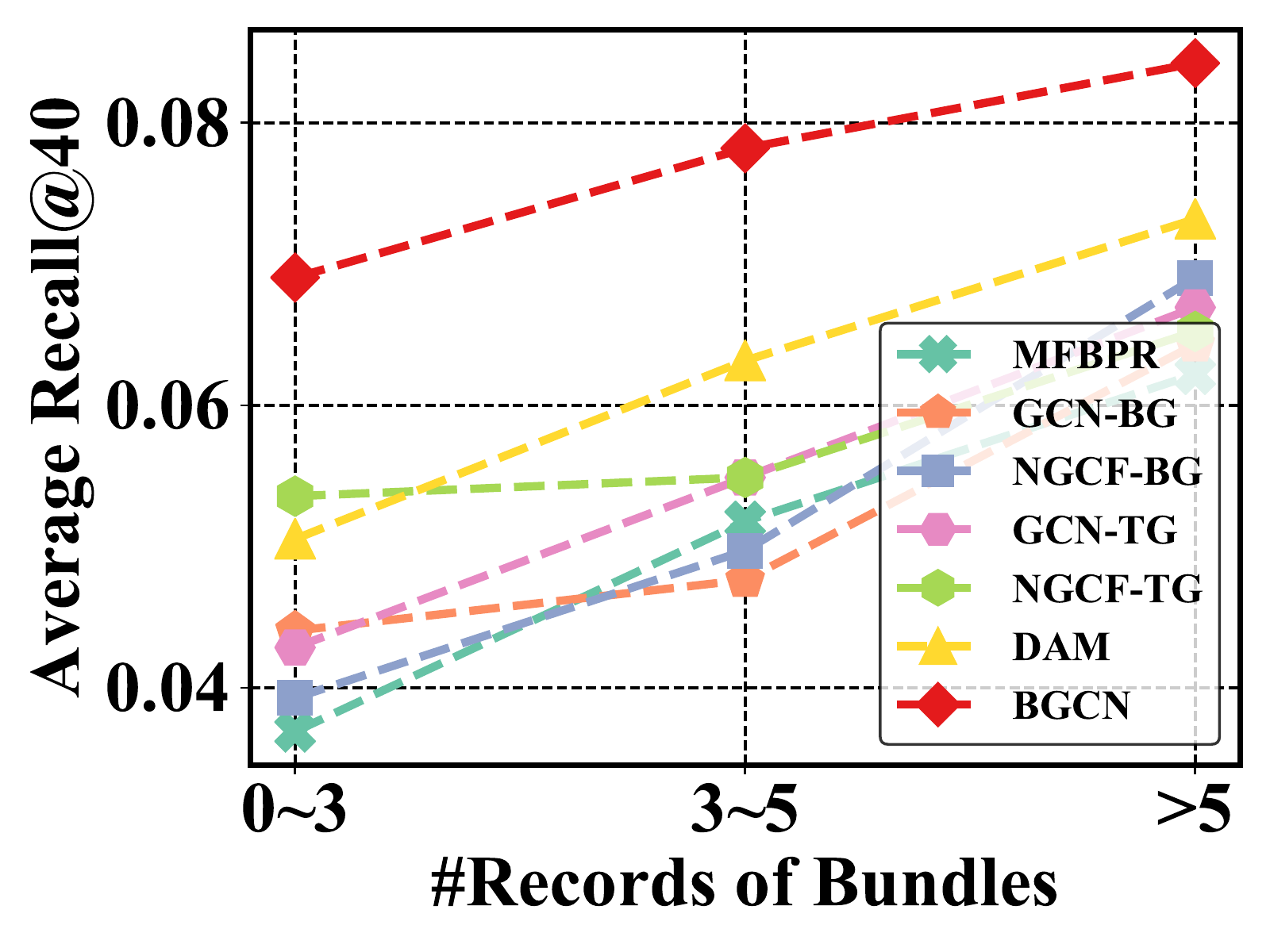}}
\hspace{0.4cm}
\subfigure
{\includegraphics[height=2.7cm]{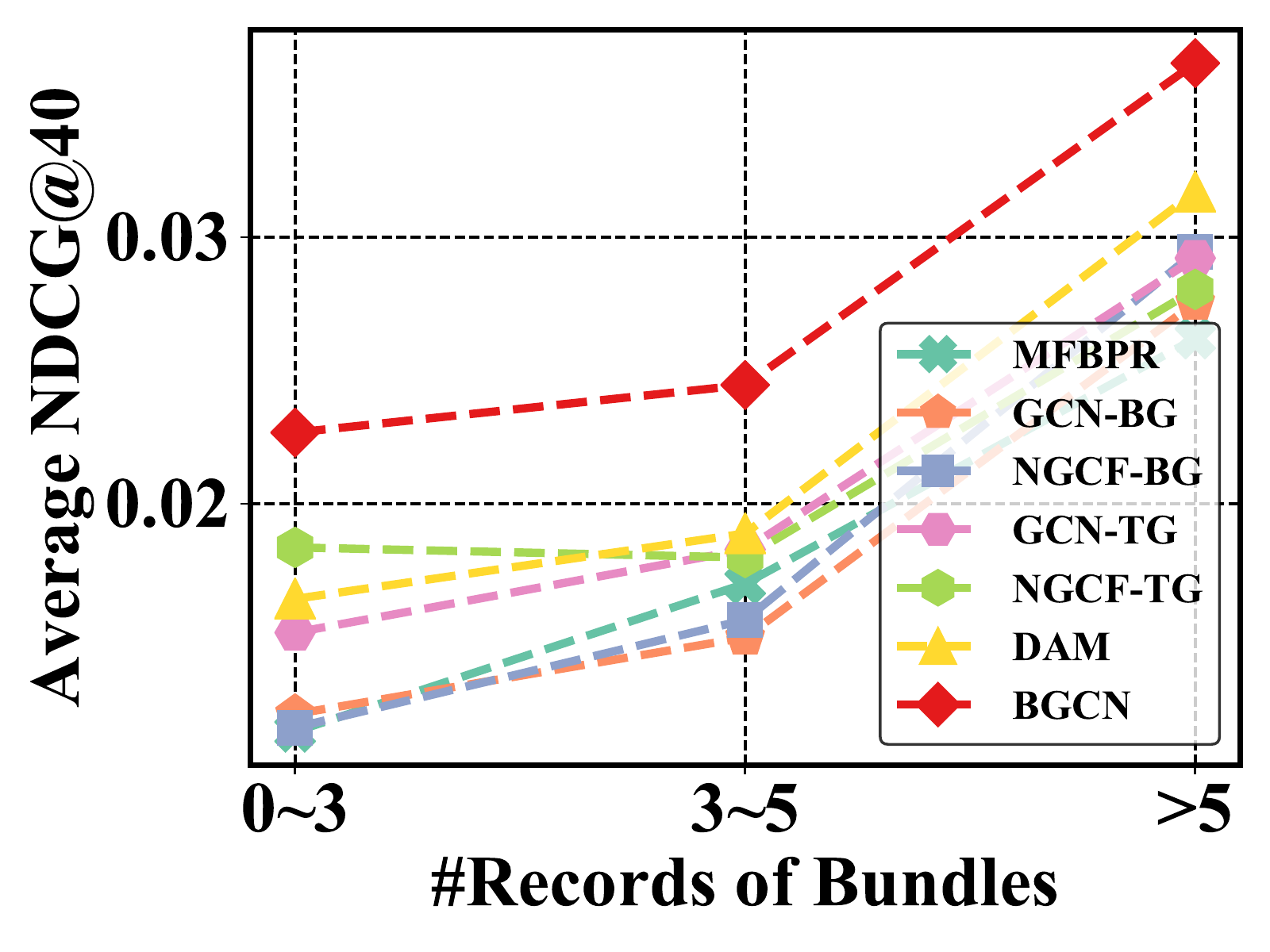}}
\vspace{-0.5cm}
\caption{Performance with different data sparsity
}\label{fig::sparsity}
\vspace{-0.5cm}
\end{figure}

\vspace{-0.1cm}
Compared to items, bundles have extremely rare interactions, 
so data sparsity is a significant issue for bundle recommendation.
To study whether our method can alleviate the data sparsity issue, 
we divide users into three groups according to sparsity as shown in Figure~\ref{fig::sparsity}
and present the recommendation performance of the Netease dataset. 
From the results, we can observe that the performance gap between BGCN and other methods becomes larger when data become sparser,
except for the NGCF-TG which is very similar to our model.
Furthermore, even in the user group with only 0\textasciitilde3 purchase records,
BGCN still keeps an excellent recommendation performance of 0.069 for Recall@40 and 0.023 for NDCG@40,
which outperforms the best method NGCF-TG by 30.19\% and 27.78\%, respectively.
As a summary, we conclude that our proposed BGCN model alleviate data sparsity issue efficiently to some extent.

\begin{table}[t]
\footnotesize
\centering
\vspace{-0.1cm}
\caption{Ablation study of the key designs}
\label{tab::ablation}
\vspace{-0.3cm}
\setlength{\tabcolsep}{0.5mm}{\begin{tabular}{llcccccc}
\toprule
\multicolumn{2}{c}{\multirow{3}{*}{\textBF{Model}}}
& &\multicolumn{2}{c}{\textBF{Netease}} & &\multicolumn{2}{c}{\textBF{Youshu}} \\
\cmidrule(lr){4-5}  \cmidrule(lr){6-8} 
&
&
& \textBF{Recall@40}   & \textBF{NDCG@40}
&
& \textBF{Recall@40}   & \textBF{NDCG@40}\\
\midrule
\multirow{3}{*}{\makecell[l]{Levels \\ Propagation}
} 
  & Item Level
&  & 0.0121 & 0.0046 &  & 0.0786 & 0.0419 \\
& Bundle Level
&  & 0.0685 & 0.0284 &  & 0.2805 & 0.1387 \\
& I\&B Levels
&  & 0.0749 & 0.0317 &  & 0.3124 & 0.1425 \\
\midrule
\multirow{3}{*}{\makecell[l]{B2B \\ Propagation}
} 
& No B2B
&  & 0.0708 & 0.0297 &  & 0.2866 & 0.1400 \\
 &Unweighted B2B
&  & 0.0738 & 0.0312 &  & 0.3040 & 0.1418 \\
& Weighted B2B 
&  & 0.0749 & 0.0317 &  & 0.3124 & 0.1425 \\
\midrule
\multirow{4}{*}{\makecell[l]{Hard-negative \\ Sample}
} 
& No Hard 
&  & 0.0749 & 0.0317 &  & 0.3124 & 0.1425 \\
& Item Level
&  & 0.0807 & 0.0343 &  & 0.3235 & 0.1573 \\
& Bundle Level 
&  & 0.0816 & 0.0343 &  & 0.3240 & 0.1581 \\
& I\&B Levels 
&  & 0.0829 & 0.0346 &  & 0.3248 & 0.1593 \\
\bottomrule            
\end{tabular}
}
\vspace{-0.4cm}
\end{table}

\section{Related Work}
\vspace{-0.1cm}
Although bundles are currently widely used everywhere, few efforts have been made in solving the bundle recommendation problem.
List Recommendation Model (LIRE)~\cite{liu2014recommending} and Embedding Factorization Machine (EFM)~\cite{EFM} 
simultaneously utilized the users' interactions with both items and bundles under the BPR framework. 
The Bundle BPR (BBPR) Model~\cite{pathak2017generating} made use of the parameters previously learned through an item BPR model.
Recently, Deep Attentive Multi-Task Model (DAM)~\cite{DAM} jointly modeled user-bundle interactions and user-item interactions 
in a multi-task manner.

The basic idea of graph convolutional networks (GCN)~\cite{GCN} is to reduce the high-dimensional adjacency information 
of a node in the graph to a low-dimensional vector representation.
With the strong power of learning structure, GCN is widely applied in recommender systems.
Berg~\textit{et al.}~\cite{GCMC} first applied GCN to the recommendation to factorize several rating matrices.
Ying~\textit{et al.}~\cite{Pinsage} extended it to web-scale recommender systems with neighbor-sampling.
Recently, Wang~\textit{et al.}~\cite{NGCF} further approached a more general model that uses high-level connectivity 
learned by GCN to encode CF signals.

\section{Conclusions and Future Work}\label{sec::conclusion}
\vspace{-0.1cm}
In this work, we study the task of bundle recommender systems.
We propose a graph-based solution BGCN that 
re-construct the two kinds of interaction and an affiliation into the graph. 
The model utilizes the graph neural network's powerful ability 
to learn the representation of two dependent entities from complex structures.
Extensive experiments demonstrate its effectiveness on real-world datasets.
As future work, 
we plan to consider the discount factor in bundle recommendation.

\vspace{-0.2cm}
\footnotesize
\bibliographystyle{ACM-Reference-Format}
\setstretch{1.00}
\bibliography{bib_full_name}
\end{spacing}

\end{document}